\documentclass[%
 aip,prl,amsmath,amssymb,reprint,twocolumn,
]{revtex4-2}

\usepackage{graphicx}
\usepackage{dcolumn}
\usepackage{bm}
\usepackage[colorlinks=true,citecolor=blue]{hyperref}
\usepackage{tikz}
\usepackage{orcidlink}
\usepackage{array}
\usepackage{makecell}

\begin{document}

\title{Coiling instability in the kitchen}

\author{Andrzej Herczyński}
\affiliation{ 
Department of Physics, Boston College, MA 02467, USA
}
\altaffiliation{ORCID: {0000-0002-0050-276X}}
\email{andrzej@bc.edu}

\author{Maciej Lisicki}
\affiliation{ 
Institute of Theoretical Physics, Faculty of Physics, 
University of Warsaw, Pasteura 5, 02-093 Warsaw, Poland
}
\altaffiliation{ORCID: {0000-0002-6976-0281}}
\email{mklis@fuw.edu.pl}

\date{\today}

\begin{abstract}
Manipulation of viscous liquids is an essential kitchen activity – from pouring golden syrup onto a pancake to decorating a cake with whipped cream frosting, from streaming ketchup on top of French fries to dispensing molten chocolate onto a strawberry. Typical viscosities in these and many other kitchen flows, and the heights from which the streams are dispensed, make such jets susceptible to the coiling instability. Indeed, the coiling of a thin thread of poured maple syrup is a source of fascination for children and adults alike, whereas the folding of the stream of ketchup squeezed out from a plastic bottle is a phenomenon familiar to all. In this paper, we review the fluid dynamics of such kitchen flows and discuss separately the case when the substrate is stationary (honey on a toast), and when it translates (cookies on a conveyor belt) or rotates (a pancake on a spinning hot plate). It is hoped that this may encourage experimentation and enjoyment of physics in the kitchen, and perhaps even lead to more elegant if not more tasty culinary results.  
\end{abstract}

\keywords{liquid instability; coiling effect; viscous flows; gravitational jets}

\maketitle

\section{Introduction}\label{sec1:intro}

\begin{figure}
    \centering
    \includegraphics[width=\linewidth]{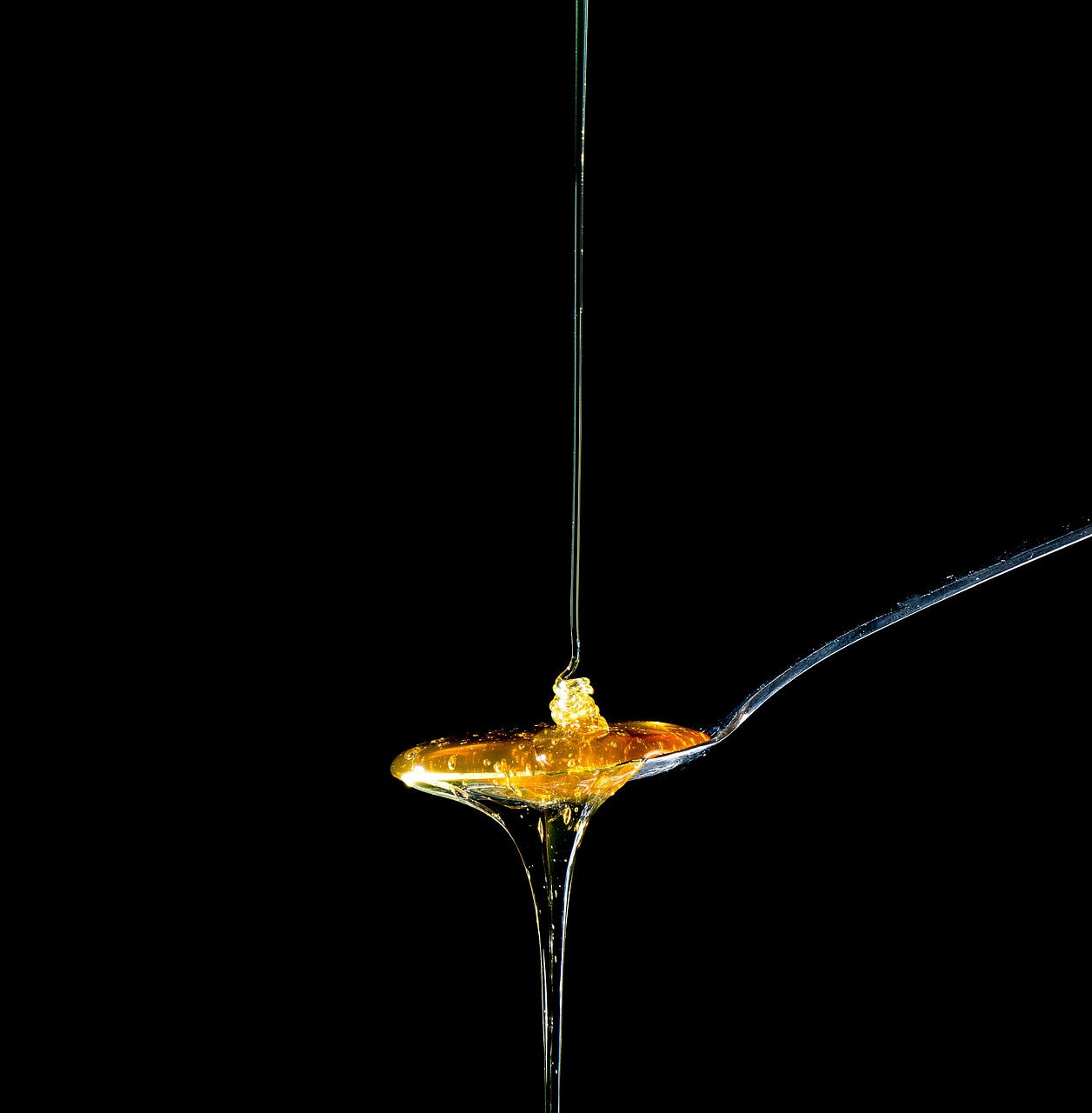}
    \caption{Coiling of honey filament on a spoon. Image by Florian Kurz from Pixabay.}
    \label{fig1}
\end{figure}

Pouring a stream of viscous liquid, such as cooking oil, ketchup, maple syrup, or honey, is perhaps one of the most ubiquitous kitchen activities – from sweetening oatmeal to decorating toasts and cookies. Viscous jets are also often created inadvertently, for example when we dip a morsel of meat in a fondue, or a strawberry in molten chocolate, and bring it over to our plate, leaving a trace on the plate, table cloth, or our shirt. In many of these flows, a coiling effect ensues, Fig. \ref{fig1}, in one of its possible forms depending primarily on the height of the liquid source above the substrate on which the jet falls, its viscosity, and flow rate, and also on the geometrical attributes of the set-up.

\begin{figure}
    \centering
    \includegraphics[width=\linewidth]{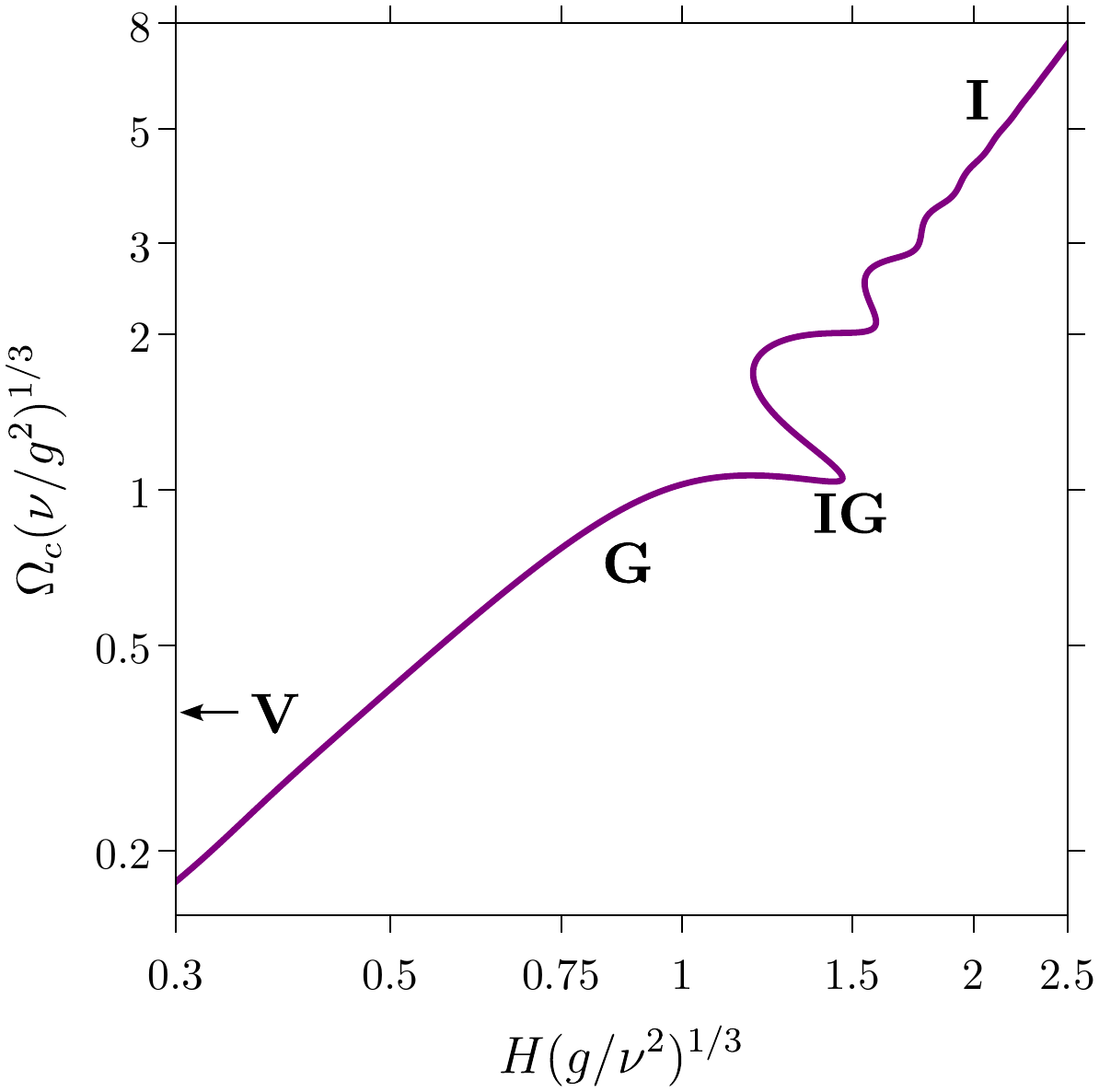}
    \caption{Dimensionless coiling frequency $\Omega{(\nu/g^2)}^{1/3}$ as a function of dimensionless fall height $H{(g/\nu^2)}^{1/3}$ for the flow rate $Q{(g/\nu^5)}^{1/3}=0.025$ indicating the coiling regimes useful for kitchen flows \cite{Lisicki2025}.   
}
    \label{fig2a}
\end{figure}

To see why coiling is frequently present in the kitchen, it is helpful to recall the four principal regimes of this instability depending on the height of fall (for a given density, viscosity, and flow rate) – viscous (V), gravitational (G), inertial-gravitational (IG), and inertial (I) \cite{Barnes1958,Mahadevan1998,Mahadevan2000,Maleki2004,Ribe2012,Ribe2017}. Consider pouring a thin stream of thick honey onto a kitchen plate. Taking the absolute viscosity of honey\cite{Munro1943} to be about $\mu=70\ \text{Pa}\cdot \text{s}$, and the typical density $\varrho\approx1.4\ \text{g}/\text{cm}^3$, we find, based on the theoretical calculations from Ref. \onlinecite{Lisicki2025} presented in Fig.~\ref{fig2a}, for a liquid of similar properties and a typical kitchen flow rate $Q\sim\ 10-100\  \text{ml}/\text{s}$, the stream will coil in the gravitational regime for heights in the range of $2 - 6$ cm, in the inertial-gravitational regime for heights in the range $6 - 12$ cm, and in the inertial regime for heights above ca. 12 cm. While the parameters for, say, pouring molten chocolate or syrup on pancakes may be different from those noted here, they will be of the same order of magnitude, and it is clear that inertial coiling is likely to occur when such a fluid is poured from a sufficient height.
 
The viscosity of many coiling liquids likely to be found in the kitchen can vary by two orders of magnitude or more, depending on the specific recipe and temperature. For example, the viscosity of maple syrup varies between 0.035 and $0.651\ \text{Pa}\cdot\text{s}$ for different grades and colors (very clear, clear, medium, amber, and dark) and temperature \cite{Ngadi2004}, with a typical viscosity\cite{Yanniotis2006} of approximately $0.164\ \text{Pa}\cdot\text{s}$ at $25^\circ\text{C}$. In contrast, the viscosity of honey is generally one order of magnitude higher than that of maple syrup and strongly dependent on the moisture (water) content, in addition to temperature. The measured honey viscosities reported \cite{Johnson1997} vary between 0.421 and $23.405\ \text{Pa}\cdot \text{s}$ for four different unifloral nectar varieties (thyme, orange, helianthus, and cotton) and may range up to $70 \ \text{Pa}\cdot \text{s}$ \cite{Munro1943}. Golden syrup, a popular replacement for honey, chilled to 12$^\circ\text{C}$, has viscosity\cite{BECKETT_2011}  $210 \ \text{Pa}\cdot \text{s}$, which rapidly decreases with temperature to ca.  $100 \ \text{Pa}\cdot \text{s}$ at room temperature\cite{Ribe2006}. It follows that if golden syrup is dispensed from a jar at sufficiently large heights of fall, about 20 cm, inertial coiling is likely to ensue for a range of flow rates, as is often observed. Indeed, children sometimes, quite intuitively, raise the jars higher above their pancakes or toasts in order to elicit the entertaining fast “swirl” of the thin thread of syrup. 

Similarly, when pouring a more viscous liquid from the kitchen cupboard, such as Heinz tomato ketchup\cite{Berta2016} with viscosity in the range of $60-160\ \text{Pa}\cdot\text{s}$, the coiling effect can occur in the gravitational regime for $H<9\ \text{cm}$, reminiscent of the way a toothpaste filament folds upon being squeezed from a tube. By contrast, for molten chocolate\cite{Lanaro2017} with viscosity around $5-15\ \text{Pa}\cdot\text{s}$, this regime occurs for $H<3\ \text{cm}$.  Other types of coiling are of course also possible, depending primarily on the liquid’s viscosity and the height of the stream. Note, however, that honey, chocolate and ketchup are non-Newtonian liquids, which alters coiling effects in a noticeable way \cite{Su2021}.

In many cases, such edible streams coil on stationary surfaces, but they may also be falling on a moving substrate, for example translating as for molten chocolate printing on a conveyor belt or rotating as in the case of oil falling on a spinning hot plate. In such cases, a great variety of patterns may be created, so the two scenarios have been dubbed fluid mechanical sewing machine (FMSM) \cite{ChiuWebster2006,Ribe2006b,Morris2008,Blount2011,Habibi2011,Herczynski2011,Brun2012,Brun2015,Ribe2022}, and its recently investigated rotational version \cite{Lisicki2022,Lisicki2025}. In what follows, we review the relevant physics in hopes of bringing attention to the assortment of coiling traces which can be easily observed in the kitchen, and encouraging experimentation, which may augment the pleasure of preparing food.

The remainder of this paper is organized as follows. In Section \ref{sec:stationary} we consider viscous streams generated in the kitchen, which fall onto a stationary surface. In Sections \ref{sec:translating} and \ref{sec:rotating} we discuss the cases where the surface is moving, either translating at a fixed linear speed or rotating at a fixed angular speed, respectively. We conclude with a few summary remarks in Section \ref{sec:conclusions}.

\section{Stationary surface}\label{sec:stationary}

Even for the simplest case, when both the source of the viscous jet and the surface on which it falls are stationary, coiling instability is a delicate, complex phenomenon, which takes different forms depending on the regime. The four distinct regimes of coiling depend on the relative magnitudes of viscous, gravitational, and inertial effects: viscous, when gravitational and inertial effects are both negligible; gravitational, when viscous and gravitational forces balance; inertial-gravitational, a multi-valued transitional regime; and inertial, when viscous forces balance liquid inertia in the coiling part of the thread\cite{Ribe2012}. For a particular fluid found in the kitchen (fixed density and viscosity), the coiling behavior depends solely on how it is poured – the flow rate $Q$ and either the radius of the thread at its origin $r_0$ or, alternatively, the speed at the top $U_0\sim Q/r_0^2$ – and the fall height $H$.

In the three stable regimes of coiling, scaling laws can be written for the frequency of coiling and other flow properties \cite{Ribe2012}. In the viscous regime, coil radius $R$ and coiling frequency $\Omega$ are proportional and inversely proportional to $H$, respectively,
\begin{equation}\label{eq:R_v}
    R_V \sim H,\qquad   \Omega_V \sim \frac{U_0}{H}.	
\end{equation}
In the gravitational regime, ignoring a multiplicative factor dependent on $H$, which may vary between 1.5 and 2 – that is with the “kitchen accuracy” – the two coiling parameters are \cite{blount_2010}  
\begin{equation}\label{eq:R_g}
    R_G\sim  \left(\frac{\nu Q}{g}\right)^{1/4},\qquad   \Omega_G\sim \frac{1}{r^2}\left(\frac{gQ^3}{\nu}\right)^{1/4},	
\end{equation}
where $r$ is the radius of the thread in the coiling tail. Finally, in the inertial regime\cite{Ribe2017,Mahadevan2000,Herczynski2011}
\begin{equation}\label{eq:R_i}
    R_I\sim \nu\left(\frac{Q}{g^2H^4}\right)^{1/3},\qquad   \Omega_I\sim \frac{1}{\nu^2}\left(\frac{H^{10}g^5}{Q}\right)^{1/3}.
\end{equation}
The radius of the coiling part of the filament, near the surface, is nearly constant in the viscous regime,
\begin{equation}\label{eq:r_v}
    r=r_0,		
\end{equation}
where $r_0$ is the radius of the thread at its origin. In the gravitational regime $r$ scales \cite{Nakata2009} as
\begin{equation}\label{eq:r_g}
    r\sim r_0\left(1+\frac{gH}{U_0^2}\right)^{-1/2}\approx r_0\left(1-\frac{gH}{2U_0^2}\right),
\end{equation}
where the last approximation holds when $gH\ll U_0^2$, which is easily satisfied in practice. Finally, in the inertial regime the thread’s thickness in the tail scales \cite{Herczynski2011} inversely with $H$,
\begin{equation}\label{eq:r_i}
    r\sim  \frac{1}{H}\left(\frac{\nu Q}{g}\right)^{1/2}.
\end{equation}

\begin{figure}
    \centering
    \includegraphics[width=\linewidth]{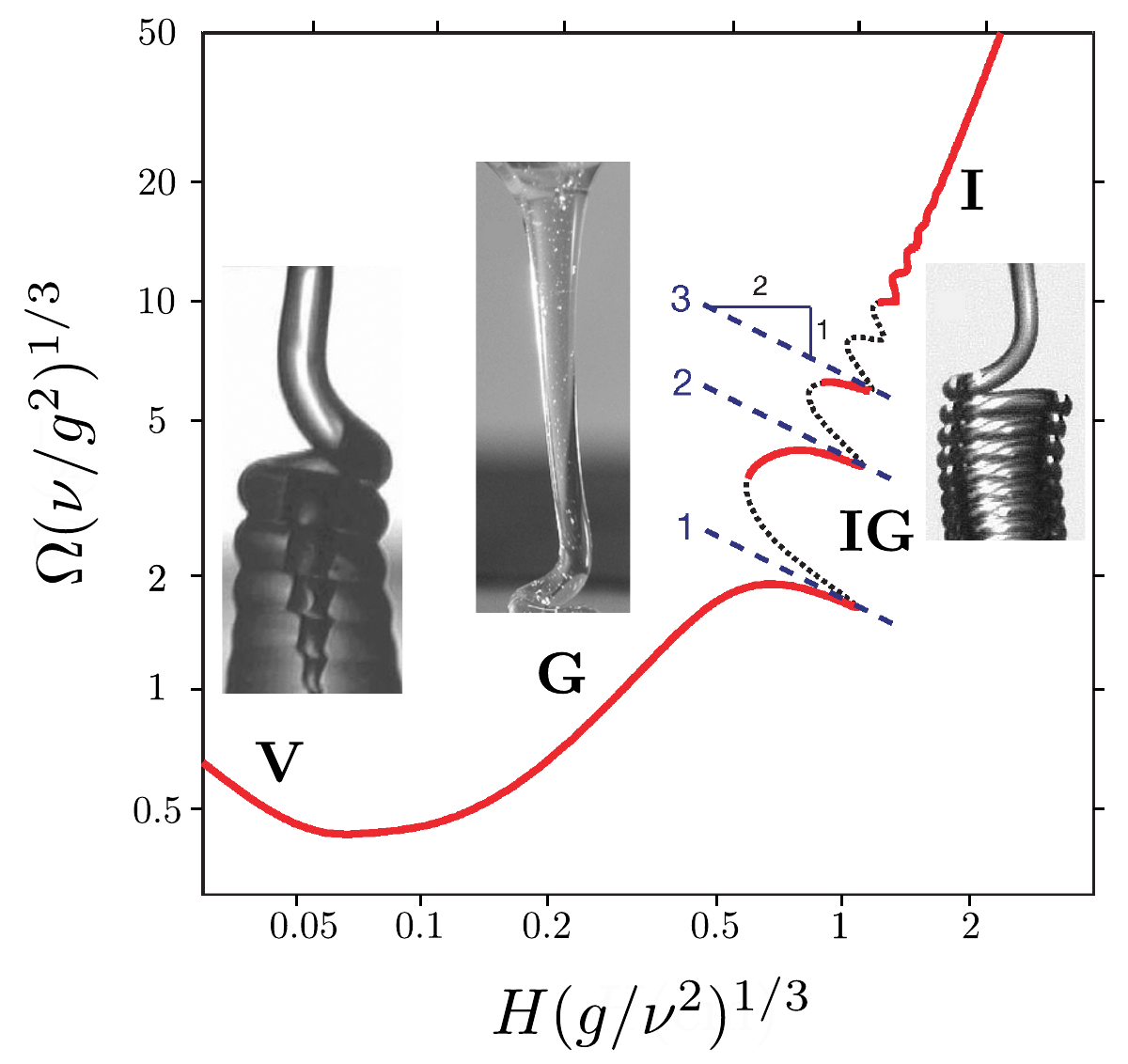}
    \caption{Dimensionless coiling frequency $\Omega{(\nu/g^2)}^{1/3}$ as a function of dimensionless fall height $H{(g/\nu^2)}^{1/3}$ for the flow rate $Q{(g/\nu^5)}^{1/3}=3.78\times 10^{-7}$ indicating the coiling regimes useful for considering kitchen flows. The dashed blue lines correspond to the “resonant” coiling frequencies, which are equal to the pendulum modes $(n=1,\ 2,\ 3\ldots)$ of the coiling tail of the thread. Adapted from Ref. \onlinecite{Ribe2022}. Photographs of coiling in various regimes (V, left; G, center; I, right) reproduced from Ref. \onlinecite{Maleki2004} with permission. 
}
    \label{fig2}
\end{figure}

The four regimes of coiling are illustrated in Fig. \ref{fig2}, which provides a plot of the nondimensional frequency of coiling $\Omega{(\nu/g^2)}^{1/3}$ as a function of the nondimensional fall height $H{(g/\nu^2)}^{1/3}$ for the dimensionless flow rate of $Q{(g/\nu^5)}^{1/3}=3.78\times 10^{-7}$.

\begin{table}[h!]
    \centering
    \begin{tabular}{|l|c|c|}
        \hline
        \textbf{Regime} & \thead{Viscous (V) or gravitational (G)}& \thead{Inertial (I)} \\ \hline
        $H(g/\nu^2)^{1/3}$ & $<1$ & $>2$ \\ \hline\hline
        \textbf{Substance} & Height [cm] & Height [cm] \\ \hline
        Chocolate\footnote{49$^\circ\text{C}$, $\mu =17\ \text{Pa}\cdot{\text{s}}$, $\varrho=1.33\ \text{g}/\text{cm}^3$} & 3 & 5 \\ 
        Honey\footnote{20$^\circ\text{C}$, $\mu =30\ \text{Pa}\cdot{\text{s}}$, $\varrho=1.4\ \text{g}/\text{cm}^3$} & 4 & 8 \\
        Golden syrup\footnote{room temperature, $\mu =100\ \text{Pa}\cdot{\text{s}}$, $\varrho=1.43\ \text{g}/\text{cm}^3$} & 8 & 16 \\
        Ketchup\footnote{23$^\circ\text{C}$,   $\mu =100\ \text{Pa}\cdot{\text{s}}$, $\varrho=1.15\ \text{g}/\text{cm}^3$ } & 9  & 18 \\
        Golden syrup\footnote{12$^\circ\text{C}$, $\mu =210\ \text{Pa}\cdot{\text{s}}$, $\varrho=1.43\ \text{g}/\text{cm}^3$} & 13 & 26 
        \\ \hline\hline
    \end{tabular}
    \caption{Typical values of fall heights associated with viscous or gravitational and inertial coiling regimes for popular kitchen fluids for the corresponding dimensionless flow rate $Q{(g/\nu^5)}^{1/3}=0.025$.}
    \label{tab1}
\end{table}

Based on Figure \ref{fig2a}, and viscosities of various liquids commonly used in the kitchen, Table \ref{tab1} lists approximate ranges of fall heights corresponding to different stable coiling regimes for chocolate, honey, ketchup, and golden syrup. The Table provides a practical guide to coiling for typical viscous liquids in the kitchen, suggesting which kind can be expected depending on the length of the stream.

It is clear that highly viscous culinary liquids, under typical kitchen conditions, may coil in the viscous mode, like a toothpaste squeezed out of a tube, with the radius of the coils growing proportionally to the height $H$ and the thickness of the thread in the coiling tail is approximately constant, (assuming constant flow rate) – as can be seen in Eqs. \eqref{eq:R_v} and \eqref{eq:r_v}. These predictions can be readily verified in the kitchen with some patient attention to such details. It may be harder to see that in this case the frequency of oscillations will decrease with the fall height, proportionally to $1/H$.

For intermediate- to high-viscosity liquids, such as very light honey, all three coiling regimes can readily be observed, with the viscous coiling accessible only for very low heights. The gravitational and inertial coiling can also be observed, with the (fairly subtle) transition from one to the other elicited simply by raising the container higher. In both these cases, the frequency of oscillations will rise with height, proportionally to $H$ for the gravitational mode, and much faster, as $H^{10/3}$, for the inertial regime, as seen in Eqs.~\eqref{eq:R_g},\eqref{eq:R_i} and \eqref{eq:r_g}.

The coil thread can be regarded as a 'machine' that converts the gravitational potential energy of the raised liquid into the kinetic energy of the spinning filament and the bending energy, with some losses due to friction. For qualitative observations, it is helpful to find how the kinetic energy per unit length of the coiling tail, $K$, depends on the flow parameters. In all three regimes, $K \sim \varrho r^2R^2\Omega^2$. With the scaling $\Omega\sim U/R \sim Q/Rr^2$, we find that $K\sim {\varrho Q}^2/r^2$. Using Eqs.~\eqref{eq:R_v}-\eqref{eq:r_i}, the kinetic energy per unit length of the thread in the three regimes can then be written as
\begin{align}
    K_V&\sim \frac{\varrho Q^2}{r_0^2},	\label{eq:k_v}						\\
K_G &\sim \varrho gr_0^2 H,		\label{eq:k_g}					\\
K_I &\sim \frac{\varrho gQ}{\nu}H^2.	\label{eq:k_i}
\end{align}
Kinetic energy in the viscous regime, Eq.~\eqref{eq:k_v}, does not depend on the fall height $H$ nor gravity as inertial and gravitational effects can be neglected in this case, nor explicitly on the viscosity. However, viscosity effectively enters via the flow rate $Q$. In the gravitational regime, wherein viscous and gravitational forces balance, kinetic energy depends only on the initial potential energy $\varrho g H$ and $r_0$, Eq.~\eqref{eq:k_g}. By contrast, the kinetic energy in Eq.~\eqref{eq:k_i}, for the inertial regime, depends on all three, viscosity, gravity, and the fall height.

There is one more aspect worth commenting on in connection with viscous coiling on a stationary surface in the kitchen – the question of the direction of the spin. In principle, the tail of the thread can coil in either direction via a spontaneous symmetry-breaking process, a classical analogue of the Goldstone mechanism in quantum mechanics. Which direction is selected depends subtly on the precise details of the initial contact of the falling thread with the surface – an intriguing aspect inviting an extended investigation in the kitchen!

\section{Translating surface}\label{sec:translating}

Although it is a common practice to pour viscous liquid onto a stationary surface while moving the vessel containing it across, we limit the discussion in this section to the opposite case – wherein the stream is stationary but the surface onto which it falls translates at a constant speed.

It may be tempting to consider the two scenarios as equivalent, differing only in the choice of the reference frame. However, the physics in the two situations is not exactly the same. In the first instance, the thread is laid along a stationary surface being pulled from the top, whereas in the second instance it is dragged by a moving surface from below and must accommodate to the surface velocity. Nevertheless, the difference in the physics is rather subtle and the resulting patterns are expected to be quite similar. Any differences between the two cases are not likely to be noticeable in observations made in the kitchen. 

It should also be noted that the former scenario has yet to receive attention in the literature, perhaps because experiments would be challenging to conduct. We therefore focus here on the case when the source of the viscous stream is stationary and the surface translates.

This scenario, the fluid mechanical sewing machine (FMSM), first described two decades ago by Sunny Chiu-Webster and John Lister \cite{ChiuWebster2006}, has now been explored both experimentally and theoretically \cite{Ribe2006b,Morris2008,Blount2011,Habibi2011,Herczynski2011,Brun2012,Brun2015,Ribe2022,Nakata2009,sze2024}. In the experiments, a viscous thread falls onto a moving belt creating a rich variety of stable “stitching” patterns depending on fluid properties, the height of fall, and, crucially, the speed of the belt. In addition, a plethora of unstable and transitory patterns may be observed, particularly in transition from one of the regimes to another, as explored for an elastic thread\cite{Habibi2011}. The transient effects have not yet been fully described in the literature. Neither has the nomenclature, even for the stable stitching shapes, been standardized. This is at least in part because some of these patterns appear in a variety of subtly different forms, and rarely all are present in any particular experiment.

\begin{figure}
    \centering
    \includegraphics[width=\linewidth]{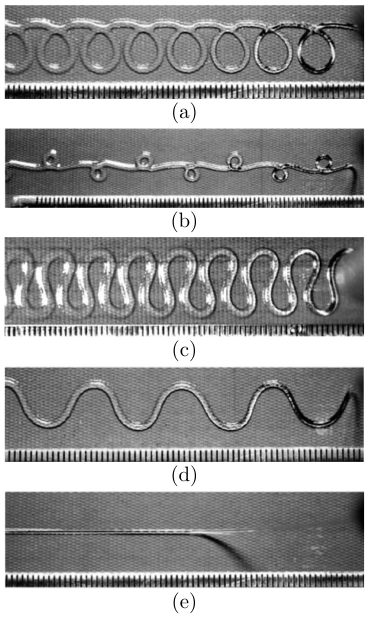}
    \caption{The “main sequence” of stitch patterns in the FMSM at increasing speed of the belt. (a) coils; (b) “one-by-one" (c) bunched-up meanders or “braiding"; (d) meanders; (e) catenary. The photographs are taken from Ref. \onlinecite{ChiuWebster2006}, courtesy of J. R. Lister.}
    \label{fig4}
\end{figure}
For all the complexity, there are four patterns that appear regularly (possibly in somewhat distinct variants) and at all fall heights $H$ in the gravitational, gravitational-inertial, and inertial regimes, evolving as the speed of the surface increases: stretched coils, Fig. \ref{fig4}(a); “one-by-one" pattern, Fig \ref{fig4}(b); meanders (possibly slanted), Fig. \ref{fig4}(c) and (d); and catenary, Fig. \ref{fig4}(e). We refer to these series as the “main sequence” of the FMSM.

\begin{figure*}
    \centering
    \includegraphics[width=\textwidth]{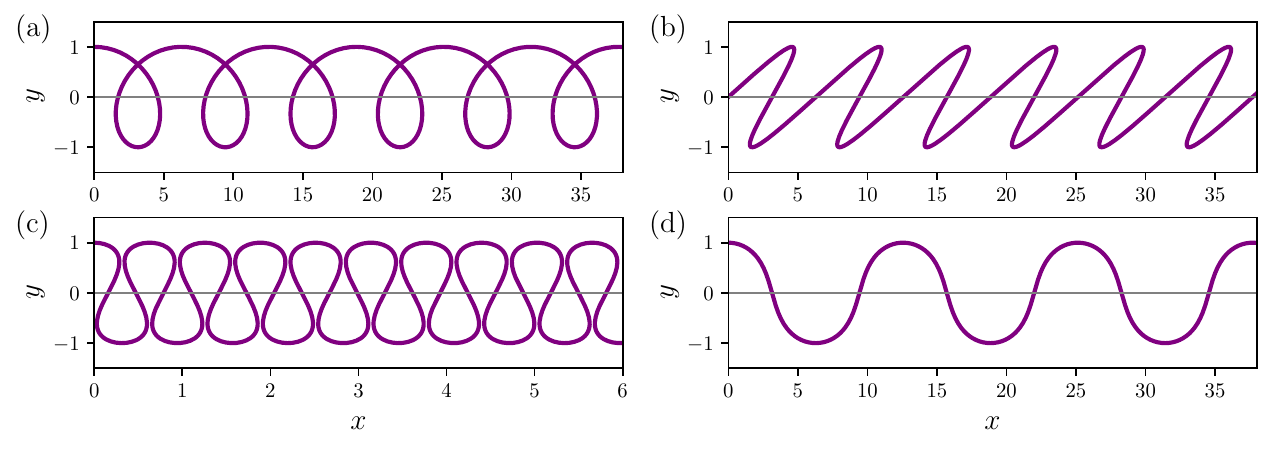}
    \caption{Parametric plots superposing oscillations and translations in the
longitudinal direction ($x$) with oscillations in the transverse direction ($y$).
(a) translated coils, $x=t+3\sin t,\ y=\cos t$; (b) slanted loops, $x=t+3\sin t,\ y=\sin t$; (c) bunched-up meanders $x=0.1t+0.2\sin 2t,\ y=\cos t$; (d) meanders
$x=t+0.5\sin t,\ y=\cos 0.5t.$
}
    \label{fig5}
\end{figure*}

All patterns observed in the FMSM experiments have been reproduced in full numerical simulations\cite{Brun2012}, and many of them also in a reduced “geometrical model” devised by Pierre-Thomas Brun {\it et al.} \cite{Brun2012,Brun2015}. A simple, qualitative realization of the stitching forms can also be obtained by superposing transverse oscillations with longitudinal translations and oscillations\cite{sze2024}. Figure \ref{fig5} displays a few examples.

Admittedly, conveyor belts are not standard equipment in domestic kitchens, although they are commonly used in decorating cakes and cookies with frosting in automated production facilities. Culinary experiments with fluid dynamical stitching are thus mostly limited to the patterns created by a coiling stream of fluid moved laterally above a substrate.

It is worth noting that such a process is akin to the painting technique developed by Jackson Pollock, an American abstract expressionist artist who painted on horizontally stretched canvases and paper. Pollock had in fact in some of his works created similar patterns to those shown in Fig. \ref{fig4} by letting a stream of highly viscous enamel paint fall on paper from sufficient height (about 20 cm or more) to elicit coiling while moving his hand laterally \cite{Herczynski2011,Cernuschi2008}. The artistic possibilities in the kitchen, inspired by Pollock’s work, seem unlimited!

\section{Rotating surface}\label{sec:rotating}

The rotational version of the fluid dynamical sewing machine (R-FMSM), whereby the viscous filament falls on a spinning surface, has only recently been investigated experimentally and analyzed theoretically \cite{Lisicki2022,Lisicki2025}, yet it is not uncommon in the kitchen and, arguably, easier to observe and explore than the FMSM.

Perhaps the ideal case is provided by the customary way of making the Chinese Shangdong pancake, which is fried on a large spinning circulate plate, about 40 cm in diameter \cite{Lisicki2022}. While the crepe is spinning, viscous syrup may be streamed on it from above. But other, more common devices, at least in the Western culinary tradition, can easily be adopted for experimenting with the R-FMSM in the kitchen, such as electric rotating cookers, available in many versions, and also stir-fry cooking vessels, which are magnetically mounted on spinning bases. Finally, the simplest possibility of all, some skillets can be rotated by hand using a vertical grip the better to mix the cooking ingredients – so one can dispense a stream of culinary fluid while simultaneously vigorously turning the pan.

All of the trace patterns obtainable in the FMSM, the main sequence among them, can still be observed in the rotational case, but they will be altered by the loss of transverse symmetry and centrifugal effects. Spinning the surface expands the manifold of possible patterns. In particular, translated coils may now be pointing inward, toward the center of rotation, or outward, or may even spontaneously switch from one side to the other. In such cases, the spacing between the arcs on the two sides of the trace may be different. For example, the intersecting coils pointing inwards will overlap more than those pointing out. Furthermore, even nominally symmetric patterns, such as meanders, will now be deformed due to variations in inertial effects when the dragged filament coils in the inward or outward direction. These centrifugal effects may be quite subtle but become more noticeable with diminishing radius of rotation and for small radii, say a few centimeters, may become quite prominent.

\begin{figure*}
    \centering
    \includegraphics[width=\textwidth]{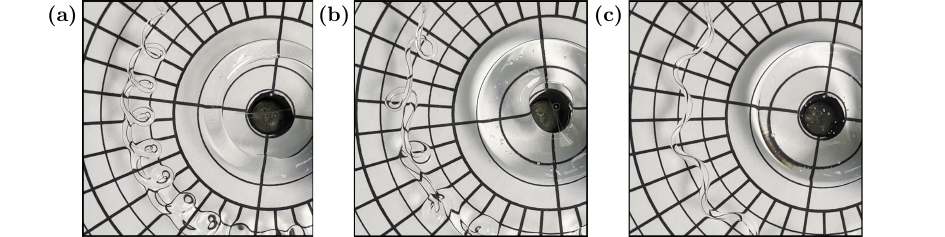}
    \caption{Asymmetric traces in the R-FMSM setup for the flow rate is $Q=2\ \text{ml}/\text{min}$, fall height $H=5\ \text{cm}$, and at the turntable radius $R=3\ \text{cm}$, at three different angular speeds. (a) translated coils; (b) “one-by-one”; (c) meanders. The working fluid is silicone oil with $\nu=0.03\ \text{m}^2/\text{s}$. Circular markings, separated by 0.5 cm, provide scale. Pictures taken from the forthcoming reference \onlinecite{Lisicki2025}. 
}
    \label{fig6}
\end{figure*}

Figure \ref{fig6} shows a few patterns in a rotating system displaying asymmetric traces in silicone oil on a turntable with a glass surface. Many other forms of rotational fluid stitching can readily be observed \cite{Lisicki2025}, including transient patterns and, particularly in the inertial-gravitational regime, disordered traces – all awaiting curiosity-driven observations or serendipitous discovery while cooking.  

\section{Concluding remarks}\label{sec:conclusions}
Coiling of a viscous thread is a common sight in the kitchen, and one of the very few fluid instabilities which are familiar broadly, and certainly to all cooks, although not often by its name (along with the Plateau-Rayleigh instability and thin film breakup). At the same time, there is something surprising, baffling even, in this phenomenon, especially when it comes to the frequency of the spin, which can reach astounding magnitudes (frequencies of over 2,000 Hz have been measured) and varies in “unexpected” ways, sometimes rising and sometimes diminishing with the rise of the fall height.

Similarly, while the radius of the thread remains nearly constant or diminishes with the height of the filament, as would be expected, the manner in which it thins out, Eqs. \eqref{eq:r_v}-\eqref{eq:r_i}, or the rate of the increase of the coil radius, Eqs. \eqref{eq:R_v}-\eqref{eq:R_i}, are complicated. It is thus seen that fluid coiling is a captivating, as well as aesthetically pleasing phenomenon, rich in possibilities. It is also ubiquitous and hard to overlook while preparing food, and can be easily appreciated by children and adults alike.

For all these reasons, coiling provides an inviting opportunity for experimentation in the kitchen, a natural bridge from culinary pursuits to explorations of physics. Are experts in fluid mechanics better cooks because of their training? They are likely to handle culinary liquids more deftly, but may also become overly distracted by the beautiful phenomena unfolding while they do so\cite{Mathijssen2023}. In any case, it may enhance the pleasure of handling culinary liquids for anyone to keep in mind that, as the physicist Peter Barham notes \cite{Barham2013}, “the kitchen is a laboratory, and cooking is an experimental science.”

\begin{acknowledgments}
This contribution would not have been possible without help from many colleagues whose research paved the way in this fascinating field. The authors are especially grateful to Neil Ribe for sharing his insights, as well as numerical results. Thanks are also due to Keith Moffatt for many fruitful discussions and suggesting the rotational version of the experiment in the first place, and to John Lister for several illuminating conversations. Gratitude is also extended to Krzysztof Kempa for pointing out that coiling may be regarded as a classical manifestation of Goldstone modes.
\end{acknowledgments}

%

\end{document}